\documentclass[12pt]{article}

\makeatletter
\@addtoreset{equation}{section}

\makeatother

\begin{document}
\setlength{\unitlength}{0.2cm}

\title{Universality of subleading exponents induced by one dimensional defects:
      the case of self-avoiding walks  }
\author{
  \\
  {\small Sergio Caracciolo }             \\[-0.2cm]
  {\small\it Scuola Normale Superiore and INFN -- Sezione di Pisa}  \\[-0.2cm]
  {\small\it I-56100 Pisa, ITALIA}          \\[-0.2cm]
  {\small e-mail: {\tt Sergio.Caracciolo@sns.it}}     \\[-0.2cm]
  \\[-0.1cm]  \and
  {\small Maria Serena Causo}             \\[-0.2cm]
  {\small\it Dipartimento di Fisica and INFN -- Sezione di Lecce}  \\[-0.2cm]
  {\small\it Universit\`a degli Studi di Lecce}        \\[-0.2cm]
  {\small\it I-73100 Lecce, ITALIA}          \\[-0.2cm]
  {\small e-mail: {\tt causo@le.infn.it}}     \\[-0.2cm]
  \\[-0.1cm]  \and
  {\small Andrea Pelissetto}                          \\[-0.2cm]
  {\small\it Dipartimento di Fisica and INFN -- Sezione di Pisa}    \\[-0.2cm]
  {\small\it Universit\`a degli Studi di Pisa}        \\[-0.2cm]
  {\small\it I-56100 Pisa , ITALIA}          \\[-0.2cm]
  {\small e-mail: {\tt pelisset@ibmth.difi.unipi.it}}   \\[-0.2cm]
  {\protect\makebox[5in]{\quad}}  
  \\
}
\vspace{0.5cm}
\date{}
\maketitle
\thispagestyle{empty}   

\vspace{0.2cm}

\newcommand{\be}{\begin{equation}}
\newcommand{\ee}{\end{equation}}
\newcommand{\<}{\langle}
\renewcommand{\>}{\rangle}

\newcommand{\R}{\hbox{{\rm I}\kern-.2em\hbox{\rm R}}}

\newcommand{\reff}[1]{(\ref{#1})}

\section{Introduction}

The presence of operators which are called {\em irrelevant}, in the terminology
of the renormalization group approach, do not change the values of leading
critical exponents. 

The determination of subleading exponents is generally a hard task. 
But there are systems, an example is studied in~\cite{Gio}, which are simpler
to study: in these cases one can find observables for which an irrelevant operator
contribute at the leading order. This happens when the irrelevant
operator breaks some symmetries of the system. In such a situation, 
a generic observable, which is not invariant under the action of the broken symmetry,
has a non-vanishing mean value, whose scaling behaviour is controlled by a
correction-to-scaling exponent.

The problem considered in~\cite{Gio} is that of a
self-avoiding-walk (SAW) which must avoid a semi-infinite 
needle, corresponding for
instance to the positive $z$-axis. This extra interaction breaks the inversion
symmetry 
$z \to -z'$. As a consequence, when this interaction is irrelevant, 
in the critical limit $N \to
\infty$, where $N$ is the length of the walk, $\<z\>$ ($z$ being the component of
the  end-point) will  scale as   $N^{\nu - \Delta}$, where $\nu$ is the universal  
critical exponent for the correlation length, and $\Delta$ is the
subleading exponent, that ``a priori" depends on the specific
symmetry broken by the needle.

In this paper we offer some simple and quite general arguments which
suggest that the first subleading exponent $\Delta$ does not
depend on the set of broken symmetries, but only on the dimensionality
of the excluded region. An explicit value for this exponent is conjectured. 
We reserve
analytical and numerical details to a  forthcoming paper.

\section{Dimensionality of defects and correction
to scaling}
We intuitively expect that the asymptotic critical
behavior of a random object $W$, of dimension $d(W)$,
embedded in a space of dimension $d$, is not
modified by the exclusion of a set $V$, of dimension $d(V)$, when 
the probability of intersection of these two objects
is vanishing. This depends only on the dimensionality of the involved 
geometrical objects.
The fundamental rule in geometric probability for 
the dimension of the generic intersection
$V\cap W$ of the two geometric sets $V$ and $W$ is 
\be
d(V\cap W) = d(V) + d(W) - d \label{dim}
\ee
If this quantity becomes negative, then $V$ and $W$ do not
generically intersect in regions outside any bounded volume in the space
$d$.
Equation~\reff{dim} holds also for a random geometry, where
one considers a probability measure on a configuration space
which is concentrated on a set of events with given Hausdorff
dimension. 
An interesting application in this context is the 
well-known random-walk representation of the
euclidean $O(n)$-vector field theory~\cite{Symanzik-V,FFSbook}.
In this representation, the interaction is concentrated on the intersections
of walks. As the generic random walk has 
Hausdorff dimension 2, for $d>4$ in the critical region
only a trivial theory is recovered because the dimension of the
generic intersection is negative.
\begin{itemize}
\item In a generic space of dimension $d>4$, this probability scales
towards a limiting value with the correlation-length as $\xi^{4-d}$,
i.e. as $N^{\nu \left(4-d\right)}$, and critical exponents have
their free-field values, coinciding with the mean-field ones. The interaction
term is an {\em irrelevant} operator in the Hamiltonian and acts at 
subleading order. 

\item In $d=4$ the interaction becomes {\em marginal} and is responsible
only of logarithmic corrections to the critical indices~\cite{ACF}.
It is said that 4 is the upper critical dimension of the model.

\item In $d<4$ the interaction is {\em relevant}: it changes the critical
indices and therefore the Hausdorff dimensions of the interacting
walks. 
\end{itemize}
In the same way, the models with excluded sets can be regarded
as the original model with extra-terms in the Hamiltonian.
In the case of random walks, the dimension of a generic 
intersection with the region $\cal R$ of dimension $d_{\cal R}$ is
\be
d_{int} = 2 + d_{\cal R} - d
\ee
For $d_{\cal R}=0$, the case in which a finite set of lattice sites
is excluded, 
the {\em upper critical dimension}, introducing logarithmic 
deviations, is $d=2$. For $d_{\cal R}=1$ is $d=3$, for 
$d_{\cal R}=2$ is $d=4$ and so on.

Let us now arrive at the case of SAW. Its Hausdorff dimension
is $1/\nu$ (see for example~\cite[Appendix B]{ACF}) and the
dimension of a generic 
intersection with the region $\cal R$ is
\be
d_{int} = {1\over \nu} + d_{\cal R} - d
\ee
Using the Flory formula for $d\leq 4$,
$\nu\approx 3/(d+2)$, which is exact in $d=1,2$, must be corrected
by logarithmic violations in $d=4$, and is a good approximation in
$d=3$, we can say that excluded points and related problems as the
 persistence problem, introduce relevant perturbations 
if $d\leq 1/\nu$, i.e. $d=1$. Excluded infinite needles are relevant 
perturbations in $d=1,2$, but already irrelevant in $d=3$, so we
expect that the probability of intersection scales towards
a limiting value as $\xi^{2-1/\nu}$, i.e. as $N^{2\nu -1}$.

It follows that 
 the average values $\<x_i^k\>_N$
of the $k$-th power of the $i$-th component of the coordinates of 
the end point of the walk in the ensemble at fixed number $N$ of
steps behave as
\be
\< x_i^k \>_N = N^{k \nu} \left( A(k) + {B_{i,{\cal R}}(k)\over
N^\Delta } + \cdots \right) \label{coordinate}
\ee
where the subleading exponent $\Delta$ is given according to our
previous discussion by 
\be
\Delta = -d_{int}\nu = 2 \nu -1.
\label{exponent}
\ee
Using the estimated value $\nu=0.5877\pm 0.0006$~\cite{Sokal2}, ~\reff{exponent} gives
$\Delta=0.1754 \pm 0.0012$.
There is a strict connection between criticality in SAW, which provides a model
for polymers in a good solvent~\cite{PJ}, and fixed-point
Hamiltonian in the 
 $O(n)$
$\sigma$-model analytically continued to 
$n=0$~\cite{Daoud,DC,ACF,FFSbook}.
We can exploit this analogy looking at
the critical behavior of SAW's in terms of the scaling form of
the correlation function between a spin at the origin and one in the bulk
at location $\vec{r}$
\be
G_{\cal R}(\vec{r};\beta)  \;\sim\;
   r^{- (d-2+\eta_{\cal R})} F_{\cal R}(\vec{r}/ \xi(\beta)) \,+\,
   r^{- (d-2+\eta'_{\cal R})} F'_{\cal R}(\vec{r}/ \xi(\beta)) \,+\,
      \ldots   \;.
 \label{GI}
\ee
and regarding the excluded regions as vacancies in the $O(n)$
$\sigma$-model.
In eq.~\reff{GI} $\beta$ is the inverse temperature, and
$\xi \sim (\beta_c-\beta)^{-\nu}$ is the correlation length
{\em in the unperturbed theory}\/;
the critical inverse temperature $\beta_c$ and
the exponent $\nu$ are not modified
by the presence of the vacancies,
unless the excluded region ${\cal R}$ is so big
that  ${\bf Z}^d \setminus {\cal R}$ 
has lower dimensionality.
If the perturbation is irrelevant, the leading exponent $\eta_{\cal R}$
and the scaling function $F_{\cal R}$ are the
same as the ones in the bulk $\eta$ and $F$.  
   The presence of the
vacancies appears in the subleading exponent $\eta'_{\cal R}$ and in the functional
form of $F'_{\cal R}$. Notice that $F_{\cal R}$ is isotropic, while  $F'_{\cal
R}$ will have a non-trivial angular dependence.
Let the observable $O$ be a function of the end-point
position. Its mean value is given by
\be
\langle O \rangle_\beta =\frac{\int{G_{\cal R}\left(\vec{r};\beta
\right)O\left(\vec r \right)d {\vec r}}}{\int{G_{\cal R}\left(\vec{r};
\beta \right) d{\vec r}}}
\label{meanvalue}
\ee
where $G_{\cal R}\left(\vec{r};\beta \right)$ is given by~\reff{GI}. 
Now consider $O\left(\vec r \right)$ such that $O\left(g\,\vec r \right) =
-O\left(\vec r \right)$ for an element $g$ of the symmetry-group of the lattice
broken by the region $\cal R$. Of course, in~\reff{meanvalue} $F_{\cal R}$ does not
contribute and therefore the leading contribution is controlled by $F'_{\cal R}$.

\section{Models, Monte Carlo simulations and results}

\begin{figure}[t]
\setlength{\unitlength}{0.19cm}
\begin{picture}(45,55)(-20,-55)
\thicklines
\put(-10,-3){
     \begin{picture}(20,4)(-10,-3)
     \put(0,0){\circle*{0.5}}
     \put(2,0){\circle*{0.5}}
     \put(2,0){\line(1,0){8}}
     \end{picture}
     }
\put(15,-3){
           \begin{picture}(20,4)(-10,-3)
           \put(0,0){\circle*{0.5}}
           \put(2,0){\circle*{0.5}}
           \put(-2,0){\circle*{0.5}}
           \put(2,0){\line(1,0){8}}
           \put(-2,0){\line(-1,0){8}}
           \end{picture}
          }
\put(-10,-20){
           \begin{picture}(20,14)(-10,-3)
           \put(0,0){\circle*{0.5}}
           \put(2,0){\circle*{0.5}}
           \put(2,0){\line(1,0){8}}
           \put(0,2){\circle*{0.5}}
           \put(0,2){\line(0,1){8}}
           \end{picture}
           }
\put(15,-30){
          \begin{picture}(20,24)(-10,-13)
          \put(0,0){\circle*{0.5}}
          \put(2,0){\circle*{0.5}}
          \put(2,0){\line(1,0){8}}
          \put(0,2){\circle*{0.5}}
          \put(0,2){\line(0,1){8}}
          \put(-2,0){\circle*{0.5}}
          \put(-2,0){\line(-1,0){8}}
          \put(0,-2){\circle*{0.5}}
          \put(0,-2){\line(0,-1){8}}
          \end{picture}
          }
\put(-10,-47){
           \begin{picture}(20,14)(-10,-3)
           \put(0,0){\circle*{0.5}}
           \put(2,0){\circle*{0.5}}
           \put(2,0){\line(1,0){8}}
           \put(0,2){\circle*{0.5}}
           \put(0,2){\line(0,1){8}}
           \put(1.2,0.4){\circle*{0.5}}
           \put(1.2,0.4){\line(3,1){8}}
           \end{picture}
           }
\put(15,-57){
          \begin{picture}(20,24)(-10,-13)
          \put(0,0){\circle*{0.5}}
          \put(2,0){\circle*{0.5}}
          \put(2,0){\line(1,0){8}}
          \put(0,2){\circle*{0.5}}
          \put(0,2){\line(0,1){8}}
          \put(1.2,0.4){\circle*{0.5}}
          \put(1.2,0.4){\line(3,1){8}}
          \put(-1.2,-0.4){\circle*{0.5}}
          \put(-1.2,-0.4){\line(-3,-1){8}}
          \put(-2,0){\circle*{0.5}}
          \put(-2,0){\line(-1,0){8}}
          \put(0,-2){\circle*{0.5}}
          \put(0,-2){\line(0,-1){8}}
          \end{picture}
          }
\end{picture}
\caption{The six different geometries we have simulated. For each geometry
we have reported as a dot the starting point of the walk (located at the 
origin) and the starting points of the defect-needles (located at distance
two from the origin).}\label{picture}
\end{figure}
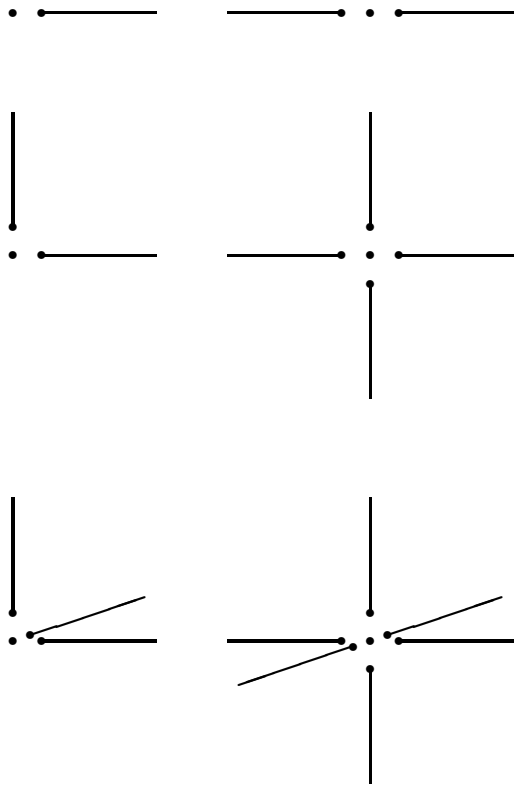

We focused our attention on the problem of the $3$--$d$ SAW on
cubic lattice, 
repelled by a region $\cal R$ which consists of a finite
collection of one dimensional regions.
The different considered geometries are shown in figure~\reff{picture}.

We performed Monte Carlo simulations for the observables that couple
only with the defects. In principle, they may show
in their leading term a symmetry dependent scaling exponent, 
in contrast with our previous claim.
We consider a plane $(y,z)$ and introduce polar
coordinates: $y = \rho \cos \theta$ and $z = \rho \sin
\theta$.  
Transformations $\theta \to \theta +
n \pi/2$ for $n$ integer),  inversion of the axis
and transformations $\theta \to \pi/2 -\theta$ are lattice symmetries.
As a consequence, suitable observables for our purposes are quantities
with angular dependence of the form:
\begin{itemize}
\item $\cos \theta$ or $\sin \theta$,
that couple with operator breaking the invariance under the $z$ and
and $y$-axis respectively,
\item $\cos 2 \theta$ or $\sin 2 \theta$, that couple with operator
 breaking the invariance under $\theta \to \pi/2 -\theta$ transformation,
\item $\cos 4 \theta$ or $\sin 4 \theta$, useful when $\cal R$ preserves 
all the lattice symmetries, as in 
the case in which it consists of six semi-infinite needles along the  
two directions of the three axis.
\end{itemize}


We performed Monte Carlo simulations on the model, using a slightly
modified version of the {\em pivot algorithm}~\cite{Sokal1}, in order to reduce
autocorrelation times for some observables.
Although the estimates were affected by systematic errors, due
to the 
neglected next-subleading terms, 
the resulting exponents were, 
for each considered ${\cal R}$, in good agreement
 with the predicted value~\reff{exponent} and confirming the conjecture  that the
leading correction to scaling  exponent, introduced in SAW by the presence of defects,
depends only on the dimension of the set of defects and {\em not} on 
the symmetries broken by this set.
Details about the Monte Carlo
simulations and results will be given in a forthcoming paper.

\end{document}